\documentclass[aps,prc,twocolumn,groupedaddress]{revtex4}

\newcommand{\beq}{\begin{equation}}
\newcommand{\eeq}{\end{equation}}
\newcommand{\beqa}{\begin{eqnarray}}
\newcommand{\eeqa}{\end{eqnarray}}
\usepackage{epsfig}

\begin{document}

 \title{The $s\bar s$ component of the proton
and the strangeness magnetic moment}

\author{B. S. Zou}
\email[]{zoubs@ihep.ac.cn} \affiliation{Institute of High Energy
Physics, CAS, P.O.Box 918, Beijing 100049, China}

\author{D. O. Riska}
\email[]{riska@pcu.helsinki.fi}
\affiliation{Helsinki Institute of Physics
and Department of Physical Sciences, POB 64,
00014 University of Helsinki, Finland}

\thispagestyle{empty}

\date{\today}
\begin{abstract}

A complete analysis is given of the implications of the
empirical indications for a positive
strangeness magnetic
moment $\mu_s$ of the proton on the possible configurations of
the $uuds\bar s$ component of the proton.
A positive value for $\mu_s$ is obtained in the
$s\bar s$ configuration where the $uuds$ subsystem is in an
orbitally excited state with $[4]_{FS}[22]_F [22]_S$
flavor-spin symmetry, which is likely to have the lowest
energy.
The configurations in which the $\bar s$ is orbitally excited,
which include the conventional
$K^+\Lambda^0$ configuration, with exception of that in which the
$uuds$ component has spin
2, yield negative values for $\mu_s$.
The
hidden strangeness analogues of recently proposed quark cluster
models for the $\theta^+$ pentaquark give differing signs
for $\mu_s$.

\end{abstract}


\maketitle

Three recent experiments on parity violation in electron- proton
scattering suggest that the strangeness magnetic moment of the
proton $\mu_s$ is positive \cite{sample,happex,a4}. In contrast theoretical
calculations have led to negative values for this observable,
with but few exceptions \cite{weigel,isgur,rho,lubov,silva,lewis}.
Here the implications of the
empirical result
for the configuration of the
constituents of the proton is considered by a calculation of
$\mu_s$ for all positive parity configurations of the
$uuds\bar s$ system with one constituent in the first orbitally
excited state, which may be contained in the
proton. The results are given in the form of simple general
expressions, in which $\mu_s$ is proportional to
the $s\bar s$ probability of the configuration.

It is shown that  $\mu_s$ is positive in the
$uuds\bar s$ configuration, which is
likely to have the lowest energy,
where the $\bar s$ quark is in the ground state
and the $uuds$ system is in the
$P-$state.
If in contrast
the strange antiquark is in the $P-$state and the 4 quarks
are in their ground state the
strangeness magnetic moment is negative (except for the
energetically unfavored case where the total spin
of the $uuds$ state equals 2).
These configurations
correspond to that of a fluctuation of the proton
into a kaon and a strange hyperon, which is well known
to lead to a negative value for the strangeness
magnetic moment \cite{musolf,forkel,hannelius,chenxs}.

Several intriguing configurations \cite{jaffe,zahed,karliner} for
the system of 4 light flavor quarks and a strange antiquarks have
been proposed to explain the structure of the (tentative)
$\theta^+$ pentaquark. Below their hidden strangeness analogues
are considered from the point of view of the strangeness magnetic
moment. While the diquark-diquark configurations suggested in
refs.\cite{jaffe,zahed} with the $\bar s$ quark in its ground
state yield positive values, the diquark-triquark configuration
suggested in ref.\cite{karliner} yields a negative value for the
strangeness magnetic moment of the proton.

If the $uuds$ quarks are in their ground state, the spatial state is
completely symmetric $[4]_X$, and the Pauli principle
requires that their flavor-spin state has the mixed symmetry
$[31]_{FS}$ so as to combine with the color state $[211]_C$ to
form the required total antisymmetry $[1111]$. The different
flavor and spin state symmetry configurations that can combine to
the required $[31]_{FS}$ mixed symmetry combination are listed in
Table 1 \cite{helminen}. The requirement of positive parity
requires that for these configurations the strange antiquark
has to be in the $P-$state.

If in contrast the strange antiquark $\bar s$ is in
its ground state, the $uuds$ system has to be in
the $P-$state in order that the combination have positive
parity. In this case the symmetry of the spatial state of
the 4-quark system is reduced to $[31]_X$. When combined
with the mixed symmetry color state $[211]_C$ this
allows the flavor-spin combination to either be
completely symmetric $[4]_{FS}$ or to have any one of the
mixed symmetries $[31]_{FS}, [22]_{FS}$ or $[211]_{FS}$.
The combinations of flavor and spin symmetry configurations
that lead to these symmetries are also listed in
Table \ref{4qsymm} \cite{helminen}.

In the quark model the strangeness magnetic moment is
defined as the matrix element of the operator
\begin{equation}
\vec \mu_s = e\sum_i {{\hat S}_i\over 2 m_s}(\vec l_i +\vec
\sigma_i)\, , \label{mus}
\end{equation}
where $\hat S$ is the strangeness counting operator,
with eigenvalue +1 for $s$ and $-1$ for $\bar s$
quarks and $m_s$ is the constituent mass of the
strange quark.
\begin{table}[hb]
\caption{Flavor and spin state symmetry configurations of
the $uuds$ quark states in the ground state and first
orbitally excited $P$ state. The states are ordered from above
after increasing matrix elements of the Casimir operator
$-\sum_{i<j}\vec\lambda_i\cdot
\vec\lambda_j\vec\sigma_i\cdot\vec\sigma_j$, where
$\lambda_i$ are the $SU(3)_F$ generators. These
matrix elements are listed in the brackets \cite{helminen}.
\label{4qsymm}}
\begin{center}
\begin{tabular}{l|l}
 $uuds$ ground state       &   $uuds$ $P-$state            \\
\hline
$[31]_{FS}[211]_F[22]_S$\, (-16)  & $[4]_{FS}[22]_F[22]_S$\,
\,\,\,\,\,\,(-28)       \\
$[31]_{FS}[211]_F[31]_S$\, (-40/3)  & $[4]_{FS}[31]_F[31]_S$\,
\,\,\,\,\,\,(-64/3)    \\
$[31]_{FS}[22]_F[31]_S$\,\,\,\, (-28/3)    & $[31]_{FS}[211]_F[22]_S$\,
(-16)     \\
%\hline
$[31]_{FS}[31]_F[22]_S$\,\,\,\, (-8)    & $[31]_{FS}[211]_F[31]_S$\,
(-40/3)    \\
$[31]_{FS}[31]_F[31]_S$\,\,\,\, (-16/3)    & $[31]_{FS}[22]_F[31]_S$\,
\,\,\,(-28/3)      \\
$[31]_{FS}[31]_F[4]_S$\,\,\,\,\,\,\, (0)     & $[31]_{FS}[31]_F[22]_S$\,
\,\,\,(-8)     \\
%\hline
$[31]_{FS}[4]_F[31]_S$\,\,\,\,\,\,\, (+8/3)     & $[4]_{FS}[4]_F[4]_S$ \,
\,\,\,\,\,\,\,\,\,\,(-8)       \\
                           & $[22]_{FS}[211]_F[31]_S $\,
(-16/3)    \\
                           & $[31]_{FS}[31]_F[31]_S $\,
\,\,\,(-16/3)    \\
%\hline
                           & $[22]_{FS}[22]_F[22]_S $\,
\,\,\,(4)   \\
                           & $[211]_{FS}[211]_F[22]_S $
(0) \\
                           & $[31]_{FS}[31]_F[4]_S $\,
\,\,\,\,\,\,(0)    \\
%\hline
                           & $[211]_{FS}[211]_F[31]_S $
(8/3) \\
                           & $[22]_{FS}[31]_F[31]_S $\,
\,\,\,(8/3)   \\
                           & $[31]_{FS}[4]_F[31]_S $\,
\,\,\,\,\,\,(8/3)     \\
%\hline
                           & $[22]_{FS}[22]_F[4]_S $\,
\,\,\,\,\,\,(4)   \\
                           & $[211]_{FS}[22]_F[31]_S$\,
(20/3) \\
                           & $[211]_{FS}[211]_F[4]_S$\,
(8) \\
%\hline
                           & $[211]_{FS}[31]_F[22]_S$\,
(8) \\
                           & $[22]_{FS}[4]_F[22]_S $\,
\quad (8)   \\
                           & $[211]_{FS}[31]_F[31]_S $\,
(32/3) \\
\hline
\end{tabular}
\end{center}

\end{table}

Consider first the configurations, in which the antiquark is in the
$P$-state. For these the spatial state of the $uuds$ is
completely symmetric $[4]_X$ with total orbital angular
momentum 0.
The corresponding wave functions for a proton with spin up may
then be written in the general form:
\newpage
\begin{eqnarray}
&&\psi =A_{s\bar s} \sum_{a,b,c}\sum_{m,s,M,J,j}\,
(1,1/2,m,s\vert J,j)\nonumber \\
&&(S,J,M,j\vert 1/2,1/2)\, C^{[1111]}_{[211]a,[31]a}
C^{[31]a}_{[F]b,[S]c}  [211]_C (a) \nonumber\\
&&[F](b)\,[S]_M(c)\, \bar Y_{1m}\,\bar \chi_s \varphi(\{r_i\})\, .
\label{sbaraa}
\end{eqnarray}
Here $[211]_C\, ,[F]$ and $[S]$ are the color, flavor and spin
states denoted by their Young patterns respectively, and the
sums over $a,b$ and $c$ run over the configurations of the
$[31][F][S]$ representations of $S_4$ for which the corresponding
Clebsch-Gordan
$C^a_{b,c}$ are non-vanishing
\cite{chen}. The orbital and spin states of the $\bar s$ are
denoted by $\bar Y_{1m}$ and $\bar\chi_s$ respectively.
$A_{s\bar s}$ is the amplitude of the $s\bar s$ component and
$\varphi(\{r_i\})$ is a symmetric function of the coordinates of
the $uuds\bar s$ system.

The configurations in the left column in Table \ref{4qsymm}
may be grouped according to their spin symmetries $[22]_S$,
$[31]_S$ and $[4]_S$. The uppermost configuration in the left
column in Table
\ref{4qsymm}, which has the spin symmetry $[22]_S$
is expected to have the lowest energy
if the $uuds$ is in its ground state. This is the case
both if the hyperfine interaction between the quarks
is described by the color magnetic interaction or
by the flavor and spin
dependent hyperfine interaction
$-C\sum_{i<j}\vec\lambda_i\cdot\vec\lambda_j\,\vec\sigma_i
\cdot\vec\sigma_j$, where $C$ is a constant with the
value $\sim 20-30$ MeV, and which leads to the empirical ordering
of the baryon resonances \cite{glozman,helminen}. For the states with
this spin symmetry the
total spin of the $uuds$ system is $S=0$. The
corresponding strangeness magnetic moment is
negative:
\begin{equation}
\mu_s = -{1\over 3}\, {m_p\over m_s}\, P_{s\bar s}\, ,
\label{sbarbb}
\end{equation}
(in units of nuclear magnetons $\mu_{_N}=e/2m_p$).
Here $P_{s\bar s}=A_{s\bar s}^2$ is the
probability of the $s\bar s$ component
and $m_s$ is the constituent mass of the strange quark.

The following two configurations in the left column in
Table \ref{4qsymm} have the spin symmetry $[31]_S$ and
thus $S=1$. For these
configurations and all other configurations in the left
column that have the same spin symmetry
$\mu_s$
may be expressed in the
general form:
\begin{equation}
\mu_s = - {m_p\over m_s} ({7\over 6} - {1\over 2}\bar \sigma)
P_{s\bar s}\, .
\label{sbarcc}
\end{equation}
Here $\bar\sigma$ is the average of $\sigma_z$ of the $s$ quark
in  the configuration for S=1 with z-projection M=1.
As $\vert\bar \sigma \vert\leq 1$, all
of these configurations yield a negative strangeness magnetic
moment in apparent conflict with the positive
empirical value. The numerical values of $\bar\sigma$ are
very similar - ie 41/72 and 1/2 - in the configurations
with flavor-spin symmetry
$[211]_F[31]_S$ and $[22]_F[31]_S$.

This result that the strangeness magnetic moment is negative if
the $\bar s$ quark is in the $P$-state generalizes to all other
configurations in the left column of Table~\ref{4qsymm} except for
the energetically unfavored
$[31]_{FS}[31]_F[4]_S$ configuration in which the $uuds$
system has total spin 2. For this configuration $\mu_s$ takes the
value:
\begin{equation}
\mu_s = {7\over 6} {m_p\over m_s} P_{s\bar s}\, .
\end{equation}

These results complete the analysis of the configurations with the
$\bar s$ quark in the $P-$state.

In the configurations $\bar s$ quark is in its ground state,
positive parity demands that the $uuds$ system has to be in
the $L=1$ state with
mixed spatial symmetry $[31]_X.$ In order to construct the
corresponding spatial wave functions one may exploit a
representation in terms of the following relative coordinates:
\begin{eqnarray}
&&\vec \xi_1 ={1\over\sqrt{2}}(\vec r_1 -\vec r_2)\, ,\quad
\vec\xi_2 = {1\over\sqrt{6}}(\vec r_1 +\vec r_2 -2 \vec r_3)\, ,
\nonumber\\
&&\vec \xi_3 ={1\over\sqrt{12}}(\vec r_1 +\vec r_2 +\vec r_3
 - 3\vec r_4)\, ,\nonumber\\
&&\vec\xi_4 = {1\over\sqrt{20}}(\vec r_1 +\vec r_2 + \vec r_3
+\vec r_4 - 4 \vec r_5)\, ,
\end{eqnarray}
which when complemented with the center-of-mass coordinate
form a complete set of basis vectors.

For these states the wave functions for the configurations
in the right columns in Table \ref{4qsymm} may for a proton
with spin up
then be written in the general form:
\begin{eqnarray}
&&\psi = A_{s\bar s}\sum_{a,b,c,d,e} \sum_{m,s,M,j}
(1,S,m,M\vert\, J,j)\nonumber\\
&&(J,1/2,j,s\vert\, 1/2,1/2)\, C^{[1111]}_{[211]a,[31]a}
C^{[31]a}_{[31]b,[FS]c}\, C^{[FS]c}_{[F]d,[S]e}
\nonumber\\
&&[211]_C(a)\,[31]_{X,m}(b)\, [F](d)\,[S]_M(e)\, \bar\chi_s\,
\varphi(\{r_i\})\, .
\label{uuds2}
\end{eqnarray}
Here $J$ is the total angular momentum of the $uuds$ system.
The three components of the spatial state may be
described as normalized combinations of a spatially
symmetric function that is multiplied by the
unit vectors $\hat \xi_1\, , \hat \xi_2$ and
$\hat \xi_3$.

The contribution to $\mu_s$ from all configurations
in the right column of Table \ref{4qsymm}, which have
the spatial symmetry $[31]_X$ and spin
symmetry $[22]_S$ are positive:
\begin{equation}
\mu_s = {m_p\over m_s}({1\over 3}+{2\over 3}\bar \ell)\,P_{s\bar s}\, .
\label{s4}
\end{equation}
Here $\bar\ell$ is the average
value of the $z-$components of the orbital angular
momentum of the $s$ quark in the $uuds$ system for
$m=1$ ($M=0$ since $S=0$ for these configurations).
These configurations thus give positive values for
$\mu_s$ as $\vert \bar \ell\vert \leq 1/2$.

For the uppermost and energetically most
likely configuration in the right column in Table \ref{4qsymm},
$\bar\ell=1/4$ and therefore $\mu_s$ is simply:
\begin{equation}
\mu_s =
{m_p\over 2 m_s}\,P_{s\bar s}\, .
\label{s1}
\end{equation}

The following symmetry configuration in the right column
Table \ref{4qsymm} has both spatial and spin
symmetry $[31]$.
The contribution to $\mu_s$ may for all states with both
spatial and spin symmetry $[31]$ be expressed as:
\begin{eqnarray}
\mu_s & =& -{m_p\over m_s}\,P_{s\bar s}\, ,\qquad\quad\quad
(J=0)\, ,\nonumber\\
\mu_s & =&  {m_p\over m_s}({1\over 3}+{1\over3} \bar \ell +
{1\over 3}\bar \sigma)\,P_{s\bar s}\, , \quad (J=1)\, . \label{s3}
\end{eqnarray}
Of these configurations that with lowest energy has
flavor-spin symmetry $[4]_{FS}[31]_F[31]_S$. In this
configuration when $J=1$ $\bar\ell = 1/4$ and $\bar \sigma = -1/18$.

>From (\ref{uuds2}) one may finally infer that the
contribution to $\mu_s$ from all the configurations
in the right column of Table \ref{4qsymm}, which have
the spatial symmetry $[31]_X$ and spin
symmetry $[4]_S$ are likewise positive
and may be
expressed in the form:
\begin{equation}
\mu_s = {m_p\over m_s}({5\over 6}-{1\over 3}\bar \ell)P_{s\bar s}\, .
\label{s5}
\end{equation}
Here the first term is the sum of the contributions
of the $s$ and $\bar s$ quarks, which are 1/2 and 1/3
respectively.
In the lowest of these configurations, which also has
flavor symmetry
$[4]_F $, $\bar \ell = 1/4$. In the following one in order of
increasing energy the flavor symmetry is $[31]_F$ and
$\bar\ell = 19/54$.
This completes the analysis of the configurations in the
right column of Table \ref{4qsymm}

These results reveal that $\mu_s$ is positive in the
lowest lying 5-quark configurations, in which the strange
antiquark $\bar s$ is in the ground state.
For the lowest configuration $\mu_s =
0.5 m_p/m_s$ times the
probability $P_{s\bar s}$ for the preformed $s\bar s$ component in
the proton. This result then leads to
$P_{s\bar s}\sim 2 m_s/m_p\,\mu_s^{exp}\sim \mu_s^{exp}$.
For $\mu_s^{exp}$ the result of the SAMPLE experiment \cite{sample}
is $\mu_s=0.37\pm 0.2\pm 0.26\pm0.07$.

In order to explain the structure of the (tentative)
$\theta^+(uudd\bar s)$ pentaquark, several clustered
quark models have been proposed
\cite{jaffe,zahed,karliner}. These quark cluster
configurations have implications also for the pentaquark
components with hidden strangeness in the nucleons. The magnetic
moments of the pentaquark states in the cluster models
have been calculated in ref.\cite{liu} for the exotic flavor
antidecuplet and normal flavor octet baryons.

In the Jaffe-Wilczek (JW) model \cite{jaffe}, the $(uuds\bar s)$
component in the
proton should have the configuration $([ud][us]\bar s)$ with two
scalar diquarks, $[ud]$ and $[us]$, in relative $P-$wave, and the
$\bar s$ quark in its ground state. The corresponding strangeness
magnetic moment is:
\begin{equation}
\mu_s ={m_p\over 3 m_s}(1+{2m_s\over m_{ud}+m_{us}}) P_{s\bar s}
\, ,
\end{equation}
where $m_{ud}$ and $m_{us}$ are diquark masses for $[ud]$ and
$[us]$, respectively.

In the Shuryak-Zahed (SZ) \cite{zahed} model, the $(uuds\bar s)$
is composed of one
scalar diquark, one tensor diquark and the $\bar s$ antiquark.
It has the
same flavor configuration as in JW model. But the orbital
excitation $\ell = 1$ is now between the two quarks inside the tensor
diquark instead of between two diquarks as in JW model. The two
quarks in the tensor diquark have total spin $S=1$, which
in combination with their relative $P-$ wave leads to
total angular
momentum $J=1$ of the tensor diquark. In this configuration
$\mu_s$ is found to be:
\begin{equation}
\mu_s = {m_p\over 2 m_s}(1+{m_q\over 3m_{us}}) P_{s\bar s}\, .
\end{equation}

Finally in the Karliner-Lipkin (KL) model
\cite{karliner}, the $\theta^+$ pentaquark is
composed of a scalar $[ud]$ diquark and a $\{ud\bar s\}$ triquark
with one orbital angular momentum $L=1$ between the diquark and
the triquark. The total spin of the triquark is one half with a
symmetric spin wave function for the two quarks.

An analogous
configuration for the $(uuds\bar s)$ component in the proton is
$[ud]\{us\bar s\}$ with $L=1$ between scalar diquark $[ud]$ and
the spin-1/2 triquark $\{ud\bar s\}$. For this configuration
$\mu_s$ is:
\begin{equation}
\mu_s = -{1\over 3}{m_p\over m_s} P_{s\bar s}\, ,
\end{equation}
which is negative definite and the same as for the $K^+\Lambda^0$
configuration.
Another analogous configuration: $[us]{ud \bar s}$ gives a positive
$\mu_s$, but with a magnitude about 100 times smaller than that
of given $[us]{ud \bar s}$. A combination of these
two configurations in the flavor SU(3) symmetry [19] still
gives a negative definite value for $\mu_s$.

These results for the quark cluster models, reveal that the
$uuds\bar s$ configurations in JW and SZ models give similar
and positive values for $\mu_s$, while the
configuration in KL model gives a negative value. A common feature
for the JW and SZ models is that in both of
these the $\bar s$ quark in
its ground state and the $uuds$ system has total angular momentum 1.
The configurations in KL and the
$K^+\Lambda^0$ meson cloud model lack that feature.

This analysis of the $(uuds\bar s)$ configurations in the
shows
that the lowest configurations with the $\bar s$ in the ground state
give positive values for $\mu_s$,
while the lowest configurations with the
$\bar s$ in the $P-$state give
negative values. The empirical indications
\cite{sample,happex,a4} for a positive $\mu_s$
implies that the $\bar s$ is in the ground state and that
the $uuds$ system is orbitally excited. This
is at variance with earlier studies on
the intrinsic strangeness in the proton based on the $K^+\Lambda^0$
configuration for the strange component
\cite{musolf,forkel,hannelius,chenxs}, which
leads to negative values for $\mu_s$.
It should be instructive to
investigate the consequences of the configurations with the
$\bar s$ in the ground state on other
processes relevant to the intrinsic strangeness, such as the NuTeV
anomaly problem \cite{NuTeV,mabq}.

Equally instructive would be a comparison of the role
of the analogous configurations for the non-strange pentaquark
components in nucleon and $N^*$ to that
of conventional meson cloud configurations. Thus
the observation of an excess $\bar d$ over $\bar u$
antiquarks in the proton \cite{garvey}, which has been described
in terms of a $n\pi^+$
configuration \cite{speth}, might alternately be
described in terms of an $[udud]\bar d$ component in the proton.
While the $N^*(1440)$ may contain significant $N\pi$ and
$\Delta\pi$
components, its unusual properties might also be described
in terms of admixtures of colored quark cluster
configurations.

A recent chiral quark model
calculation for the $uudd\bar s$
pentaquark system \cite{huang} suggests that
the configuration
with both $uudd$ and $\bar s$
in their ground state gives lower energy than
the $[31]_X[4]_{FS}$ configurations. It should be interesting to
extend that calculation to the configuration with $uudd$ in the
ground state and the $\bar s$ in its $P-$ wave state to see
if
it gives a higher energy than the lowest $[31]_X[4]_{FS}$ state.
If so, it might provide an explanation for why the
pentaquark component of the proton is in the lowest $[31]_x[4]_{FS}$
state. The penta-quark model with the most favorable
configurations of the
flavor-spin interaction with one quark in the $P-$state
\cite{stancu}  naturally leads to a positive value
for $\mu_s$.

In summary, the empirical indications for a positive
strangeness magnetic moment of the proton suggest that the $s\bar
s$ configuration in the proton is such that the
$\bar s$ is the ground state and the $uuds$ system
in the $P-$state. This suggests
that the $qqqq\bar q$ components in baryons may be
mainly in colored quark cluster configurations rather
than in ``meson cloud'' configurations.
In the configuration with the lowest
energy the positive empirical strangeness magnetic moment
gives a direct estimate of the probability of the $s\bar s$
configuration.
\vspace{0.2cm}
\begin{acknowledgments}

B. S. Zou acknowledges the hospitality of the Helsinki
Institute of Physics during course of this work. D. O. Riska
acknowledges the hospitality of the W. K. Radiation
laboratory, California Institute of Technology.
Research supported in part by the Academy of Finland grant
number 54038 and the
National Natural Science Foundation of China.

\end{acknowledgments}

\end{document}